# Miniaturized Patch Rectenna Using 3-Turn Complementary Spiral Resonator for Wireless Power Transfer

Ali Raza[1, 2], Rasool Keshavarz[1], and Negin Shariati[1, 2]

[1]RF and Communication Technologies (RFCT) Research Laboratory, University of Technology Sydney (UTS), Australia.

[2]Food Agility CRC Ltd, Sydney, Australia.

Ali.Raza-1@student.uts.edu.au, Rasool.Keshavarz@uts.edu.au, and Negin.Shariati@uts.edu.au

*Abstract*—A miniaturized linearly-polarized patch antenna is presented for Wireless Power Transfer (WPT) at $1.8\ GHz$. The proposed antenna consists of a patch element and a 3-turn Complementary Spiral Resonator (3-CSR) with antenna dimension of $50\ mm \times 50\ mm$. 3-CSR is inserted in the ground plane to reduce the antenna size. This modification also increased the impedance bandwidth from 43 MHz $(1.78 - 1.83\ GHz)$ to 310 MHz $(1.69 - 2.0\ GHz)$. Moreover, antenna is fabricated and simulated and measured results are in good agreement. Additionally, a rectifier and matching circuits are designed at $-10\ dBm$ to realize a rectenna (rectifying antenna) for WPT application. Rectenna efficiency of $53.6\ \%$ is achieved at a low input power of $-10\ dBm$.

*Keywords—Complementary spiral resonator, energy harvesting, miniaturization, patch antenna, rectenna, wireless power transfer.*

## I. Introduction

Internet-of-Thigs (IoT) systems and their applications have gained exponential development in recent years [1]. Due to this rapid growth in the number of IoT devices, wireless power transfer (WPT) has become an attractive solution to empower these devices, wirelessly [2]. WPT is a clean energy solution to wirelessly transmit RF power to mobile devices and sensor networks so they can charge themselves automatically. WPT can avoid the replacement of bulky batteries and the manual charging of short-life batteries.

RF energy harvesting can be divided into two categories: RF energy scavenging from ambient sources, and WPT using dedicated RF transmitter. In ambient RF power scavenging, energy is harvested from available RF signals (GSM, Wi-Fi) in free space [3]. In WPT scenario, a dedicated transmitter is used and the power level can be adaptive from a fixed power source. A typical RF energy harvesting system consisting of receiving antenna, matching network and rectifier (Fig. 1). Receiving antenna is responsible to capture RF power from wireless sources. The received power at point A should be transferred to point B, input of the rectifier. For maximum power transfer, a matching circuit is required to transform the input impedance of the rectifier to the 50 Ω impedance of antenna. The rectification device (e.g., diode) converts harvested AC signal to DC voltage at point C. The produced voltage at point C can be stored in supercapacitor or battery and consumed by load such as RFID, wearable biomedical sensors, or wireless sensors networks.

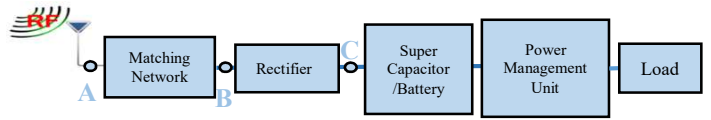

Fig. 1. Typical RF energy harvesting system block diagram (a dedicated transmitter is required in WPT scenario).

Compact antennas are desired in wireless communication systems. In such systems, physical antenna space is very limited. To this end, small antennas have been extensively studied in the past few decades and different methods have been proposed in the literature to reduce the antenna size. Meandering of line is used for size reduction where dipole antenna is folded into a cubical structure [4]. Inductive coupling is proposed for size reduction, where capacitive coupled radiating loop is fed by inductive loop for impedance matching [5]. Integration of lumped elements is adopted to reduce the size [6]. Insertion of slot in the ground plane is proposed to reduce the dimensions of patch antenna [7]. Metamaterials have also been used to miniaturize the antenna size by creating left-handed (LH) transmission line. Split ring resonator (SRR) and its dual complementary split ring resonator (CSRR) serve the purpose of size reduction [8]-[10].

Different antenna types have been reported to capture ambient RF energy including rectangle patch [11], dipole [12], and ring [13]. A single band differential antenna is presented for RF energy harvesting with board size of $137 \times 137\ mm^2$ [14] where the maximum length of element is $127\ mm$ and the antenna operates in GSM900 band with a gain of 8.5 dBi. A triple band differential antenna is presented for RF energy harvesting [15]. The antenna operates at UMTS ($2.1\ GHz$), WLAN ($2.4 - 2.48 GHz$) and WiMAX ($3.3 - 3.8 GHz$) bands with gains of 7, 5.5 and 9.2 $dBi$, respectively. The substrate size is $120 \times 120\ mm^2$ and maximum length of element is $72\ mm$. A dual-band rectenna system is presented, operating at GSM1800 and UMTS with gains of 10.9/13.3 $dBi$ at $1.85/2.15\ GHz$, respectively [16]. The substrate dimensions are $190 \times 100\ mm^2$.

In this paper, a new rectenna is proposed for WPT application at $1.8\ GHz$. A miniaturized patch antenna is designed and fabricated to receive RF power with impedance bandwidth of $1.69 - 2.0\ GHz$ (GSM1800). Miniaturization is achieved by inserting 3-turn complementary spiral resonator (3-CSR) structure on the ground plane. A rectifier circuit is also designed for RF-DC

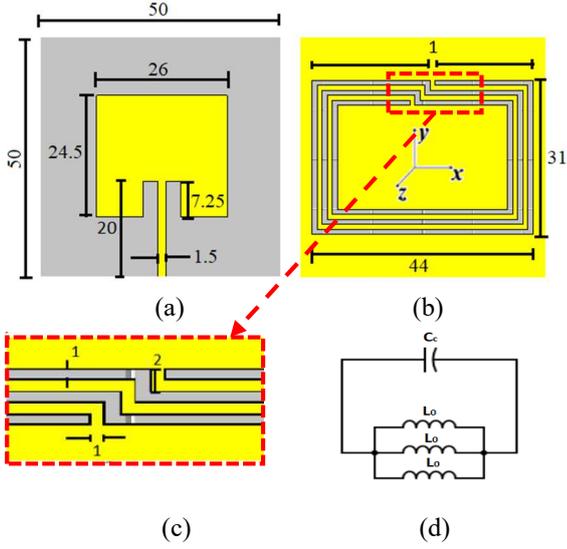

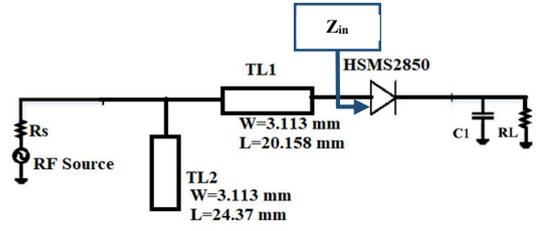

Fig. 3. Rectifier circuit with matching network at 1.8 $GHz$.

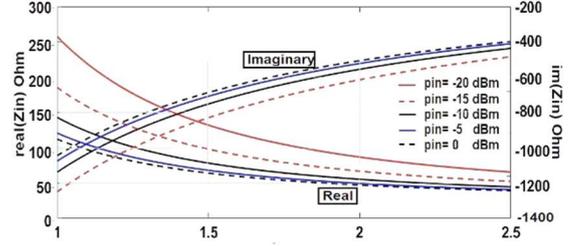

Fig. 4. Simulated rectifer input impedance over $-20\ dBm$ to $0\ dBm$ with 5 dB step.

Fig. 2. Geometry of the proposed patch antenna. (a) top view, (b) bottom view, (c) magnified view, (d) equivalent circuit model of 3-CSR structure [10], (all units are in milli-meters).

conversion at $-10\ dBm$ ($100\ \mu W$) input power.

## II. WPT SYSTEM DESIGN

This section presents antenna design in Subsection A and rectifier design in Subsection B.

### A. Antenna Design

The top and bottom sides of the proposed miniaturized antenna is shown in Fig. 2(a) and Fig. 2(b), respectively. The antenna is designed using CST MWS on commercially available FR-4 substrate with dielectric constant of 4.3, loss tangent of 0.025, and thickness of 1.6.

The goal was to design a miniaturized patch antenna for 1.8 $GHz$ (GSM1800 band). Modification is done in the ground plane of conventional patch antenna and 3-CSR structure is inserted, leading to impedance bandwidth of $1.69 - 2.0\ GHz$ with dimension of $44\ mm \times 31\ mm$ ($0.127\lambda_g \times 0.09\lambda_g$). For comparison, a conventional patch antenna at the center frequency of the proposed antenna is also simulated with patch dimensions of $39.5\ mm \times 51.16\ mm$. It is evident that size reduction is achieved by comparing conventional patch with the proposed antenna,. Additionally, impedance bandwidth increased from 43 MHz ($1.78 - 1.83\ GHz$) to 310 MHz ($1.69 - 2.0\ GHz$). The equivalent circuit model of 3-CSR structure is shown in Fig. 2(d). The inductance and capacitance values can be calculated using following equations [10].

$$L_0 = 2(L+W)L_{pul} \quad (1) \qquad L_s = \frac{L_0}{3} \quad (2)$$

$$C_c = 4\frac{\varepsilon_0}{\mu_0}L_s \quad (3) \qquad f_0 = \frac{1}{2\pi\sqrt{L_s C_s}} \quad (4)$$

Where $L_{pul}$ is the per-unit-length inductance, $L$ and $W$ is the length and width of the 3-CSR rectangle, respectively.

### B. Rectifer Design

After successful miniaturization of the patch antenna, a simple rectifier circuit is designed and simulated in ADS to create a rectenna. Rectifier converts the received RF power to DC voltage which can be used to drive a low-power load. Low turn-on voltage Schottky diode (HSMS2850) is used as a rectification device and Large Signal S-Parameters (LSSP) simulation is performed in ADS software with $-20\ to\ 0\ dBm$ input power and a load of $10\ k\Omega$. In order to match the impedance of the rectifier with the antenna impedance of $50\ \Omega$, a matching circuit is designed and simulated at $1.8\ GHz$ for $-10\ dBm$. The matching circuit along with the rectification device is shown in Fig. 3. Input impedance of the rectifier is $Z_{in} = 62.414 - j550.479\ \Omega$ at $1.8\ GHz$ for $-10\ dBm$ as shown in Fig. 4. The rectified output DC voltage and RF-DC conversion efficiency is shown in Fig. 5(a) and Fig. 5(b), respectively. As can be seen, output DC voltage at $-10\ dBm$ is 0.732 V and increasing with the input power. The maximum efficiency is $53.6\ \%$ at $-10\ dBm$, which is a suitable input power range for WPT and energy harvesting applications.

## III. RESULTS AND DISCUSSION

The proposed antenna is designed using CST MWS and fabricated on $FR-4$ substrate with dimensions of $50\ mm \times 50\ mm$ to validate simulation results. Fabricated prototype is shown in Fig. 6. Simulated and measured reflection coefficients of the proposed miniaturized antenna are in close agreement (Fig. 7). The proposed structure demonstrates a good impedance matching at 1.8 GHz. Reflection coefficient of the proposed structure is also compared with conventional patch in Fig. 7. It is clear that two advantages have been achieved using 3-CSR structure; (1) proposed antenna has resonance frequency as of conventional antenna, but with smaller dimensions of patch, and (2) impedance bandwidth has been increased.

The simulated and measured two-dimensional (2D) radiation pattern of the antenna in $xz$ and $yz$ planes are shown in Fig. 8(a) and Fig. 8(b), respectively.

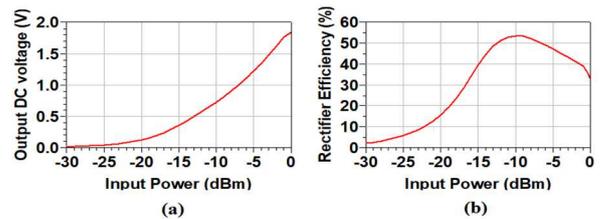

Fig. 5. Rectenna simulation results (a) output DC voltage, (b) RF-DC rectification efficiency.

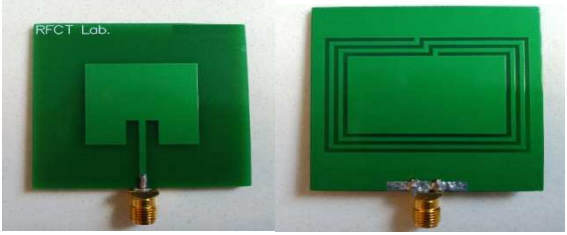

Fig. 6. Fabricated prototype of the proposed antenna.

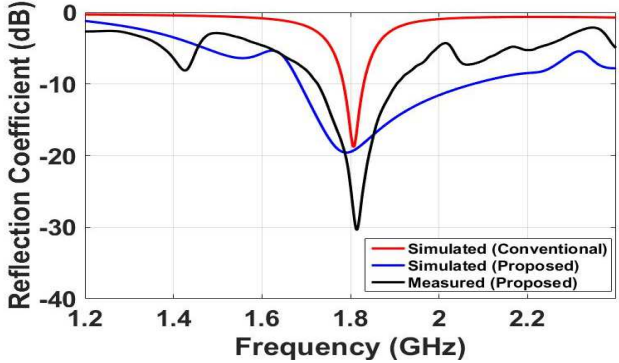

Fig. 7. Simulated and measured reflection coefficient of the antenna.

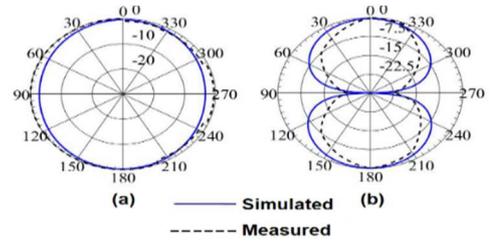

Fig. 8. Simulated and measured two-dimensional (2D) radiation patterns of the proposed antenna at 1.8 GHz.

Bi-directional pattern in $yz$ plane is due to 3-CSR structure at the bottom side of substrate. Measured and simulated results are in close agreement as shown in Fig. 8 and maximum measured gain of 2.5 $dBi$ has been achieved. A comparison of published antennas with this work is provided in Table 1. It is clear that the proposed antenna is smaller and has larger bandwidth compared to previously published papers.

Table. 1 Comparison of published antennas with the proposed structure.

| Ref | Frequency Band (GHz) | Gain (dB) | Antenna Size (mm$^2$) |
|---|---|---|---|
| [12] | $5.725 - 5.875$ | - | $18.6 \times 4.2$ ($0.24\lambda_g \times 0.052\lambda_g$) |
| [14] | $0.87 - 1.05$ | 8.5 | $127 \times 127$ ($0.195\lambda_g \times 0.195\lambda_g$) |
| [15] | $2.1, 2.4 - 2.48, 3.3 - 3.8$ | 7, 5.5, 9.2 | $72 \times 68$ ($0.29\lambda_g \times 0.23\lambda_g$) |
| [16] | $1.8 - 2.2$ | 13.3 | $190 \times 100$ ($0.825\lambda_g \times 0.44\lambda_g$) |
| Proposed Antenna | $1.69 - 2.0$ | 2.5 | $24.5 \times 26$ ($0.127\lambda_g \times 0.09\lambda_g$) |

## IV. Conclusion

A miniaturized linearly-polarized patch antenna with bi-directional radiation pattern in $yz$ plane is designed, using a 3-turn complementary spiral resonator, to receive RF power at GSM1800 ($1.69 - 2.0\ GHz$). To validate the simulation results, the proposed antenna is fabricated and measured. Additionally, a rectifier circuit along with matching network is designed and simulated at $-10\ dBm$ input power for RF-DC conversion. The rectenna achieved an efficiency of 53.6 % at $-10\ dBm$. This proves the practicality of the proposed structure for potential WPT applications.

## Acknowledgment

We would like to acknowledge the partial support of Food Agility CRC Ltd, funded under Commonwealth Government CRC Program. The CRC Program supports industry-led collaboration between industry, research and the community.